\documentclass[aps,prb,reprint,amsmath,amssymb,a4paper,citeautoscript,floatfix,showpacs]{revtex4-1} 
\usepackage{mathptmx}
\usepackage[utf8]{inputenc} 
\usepackage{graphicx}
\usepackage{todonotes}
\usepackage{xspace}
\usepackage{textcomp}
\usepackage[
,textwidth=17.5cm
,textheight=23.5cm
,verbose
,dvips
]{geometry}

\begin{document}

\title{Coupling of exciton states as the origin of their biexponential decay dynamics in GaN nanowires}

\author{Christian Hauswald}
\email{hauswald@pdi-berlin.de}
\author{Timur Flissikowski}
\author{Tobias Gotschke}
\author{Raffaella Calarco}
\author{Lutz Geelhaar}
\author{Holger T. Grahn}
\author{Oliver Brandt}

\affiliation{Paul-Drude-Institut für Festkörperelektronik,
Hausvogteiplatz 5--7, 10117 Berlin, Germany}

\begin{abstract} 
Using time-resolved photoluminescence spectroscopy, we explore the transient behavior of bound and free excitons in GaN nanowire ensembles. We investigate samples with distinct diameter distributions and show that the pronounced biexponential decay of the donor-bound exciton observed in each case is not caused by the nanowire surface. At long times, the individual exciton transitions decay with a common lifetime, which suggests a strong coupling between the corresponding exciton states. A system of non-linear rate-equations taking into account this coupling directly reproduces the experimentally observed biexponential decay. 
\end{abstract}
\pacs{71.35.-y, 
78.67.Uh,
78.55.Cr,
78.47.jd
}

\maketitle

Spontaneously formed GaN nanowires (NWs) exhibit a high structural perfection regardless of the substrate used.\cite{Geelhaar2011a}
Their geometry inhibits the propagation of dislocations along the NW axis, and the material is thus indeed virtually free of threading dislocations, which plague epitaxial GaN films.\cite{Bennett2010} Hence, it is expected that the exciton lifetimes of GaN NWs rival those measured for the highest quality epitaxial GaN layers available to date.\cite{Morkoc2001,*Scajev2012b} However, photoluminescence (PL) transients obtained for GaN NWs in time-resolved experiments do not generally exhibit a monoexponential decay as expected for a single excitonic transition. Instead, bi- and nonexponential transients were obtained,\cite{Yoo2006, Corfdir2009d, Korona2012, Gorgis2012} which impede the extraction of a single lifetime. Analogous observations were made for ZnO NWs.\cite{Wischmeier2006a, Zhao2008b} This nonexponential decay was attributed to surface-related effects by different groups.\cite{Wischmeier2006a, Zhao2008b, Corfdir2009d, Park2009, Gorgis2012} In fact, single GaN NWs with a very high surface-to-volume ratio were recently shown to exhibit individual single exponential decays,\cite{Gorgis2012} and their superposition in ensemble measurements thus inevitably results in a nonexponential transient.

In the present article, we investigate the exciton decay dynamics in GaN NWs of larger diameter. We focus on two different ordered NW arrays having narrow diameter distributions and on one spontaneously formed NW ensemble with a broad diameter distribution. The dominant radiative transition decays biexponentially for each of these samples. Neither a spectral superposition of different states nor the NW surface are responsible for these biexponential transients. Instead, we show that it is the coupling of all exciton states participating in recombination which determines their temporal evolution. This insight allows us to extract the actual lifetime of the donor-bound exciton from our experimental results. For low excitation, the values obtained are much below the radiative lifetimes of at least 1\,ns measured in free-standing GaN layers\cite{Monemar2008,Monemar2010} and are thus governed by a nonradiative decay channel that is not related to the NW surface.

\begin{figure*}[t]
\includegraphics*[width=15cm]{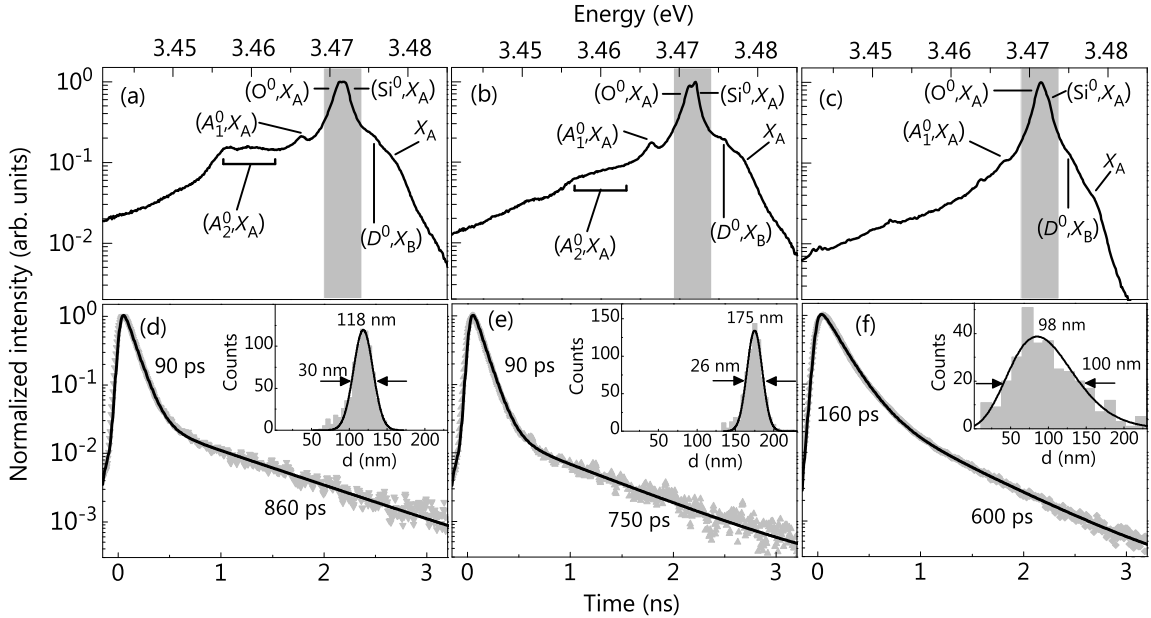} 
\caption{\label{fig:transients}Low-temperature PL spectrum and TRPL transient of sample A $\left[\text{(a) and (d)}\right]$, sample B $\left[\text{(b) and (e)}\right]$, and sample C $\left[\text{(c) and (f)}\right]$, respectively. The spectra in (a)--(c) are dominated by transitions due to donor-bound excitons [($D^0,X_{\text{A}}$)], but also acceptor-bound [($A^0,X_{\text{A}}$)] and free ($X_\text{A}$) exciton transitions are observed. The shaded areas indicate the spectral range of integration used for obtaining the TRPL transients displayed in (d)--(f). The decay times given next to the transients have been extracted by a fit (solid line) of the experimental data with a phenomenological biexponential decay convoluted with the system response function. The insets show the diameter distribution of the respective NW ensemble. The mean NW diameter and the full width at half maximum of the respective histogram (grey bars) are obtained by fits (solid lines) with a normal distribution for samples A and B and a shifted Gamma distribution for sample C.}
\end{figure*}
The three GaN NW ensembles under investigation were synthesized by plasma-assisted molecular-beam epitaxy on Si(111) substrates. Samples A and B were obtained by selective area growth (see Ref.~\onlinecite{Schumann2011b} for details regarding substrate and mask preparation) and contain spatially ordered arrays of GaN NWs with a pitch of 360\,nm and well-defined diameters of 120 and 175\,nm, respectively. Sample C is a representative example of a self-induced GaN NW ensemble (see Ref.~\onlinecite{Geelhaar2011a} for details regarding growth) characterized by a high density of NWs with random position and a broad diameter distribution with a mean of 100\,nm.  

For PL spectroscopy, the samples were cooled in a microscope cryostat to a temperature of 10\,K. In all cases, the excited area was several \textmu m in diameter and thus spanned over at least 100 NWs. Continuous-wave PL was excited by the 325\,nm (3.814\,eV) line of a He-Cd laser focused onto the samples with an excitation density of less than 1\,W/cm$^2$. The PL intensity was spectrally dispersed by a 80\,cm monochromator providing a spectral resolution of 0.25\,meV and detected with a cooled charge-coupled device array. Time-resolved (TR) PL measurements were performed by exciting the samples with the second harmonic (325\,nm) of fs pulses from an optical parametric oscillator synchronously pumped by a femtosecond Ti:sapphire laser, which itself was pumped by a frequency-doubled Nd:YVO$_4$ laser. The energy fluence per pulse was set to 0.2\,\textmu J/cm$^2$ for samples A and B and 0.8\,\textmu J/cm$^2$ for sample C. Assuming that all incident light is absorbed by the NWs, the upper limit of the photogenerated carrier density in all samples is estimated to be $5\times 10^{16}$\,cm$^{-3}$ (the higher fluence used for sample C is compensated by the higher NW density). The transient PL signal was dispersed by a monochromator providing a spectral resolution of 4\,meV and detected by a streak camera with a temporal resolution of 50\,ps.

\begin{figure*}[t]
\includegraphics*[width=15cm]{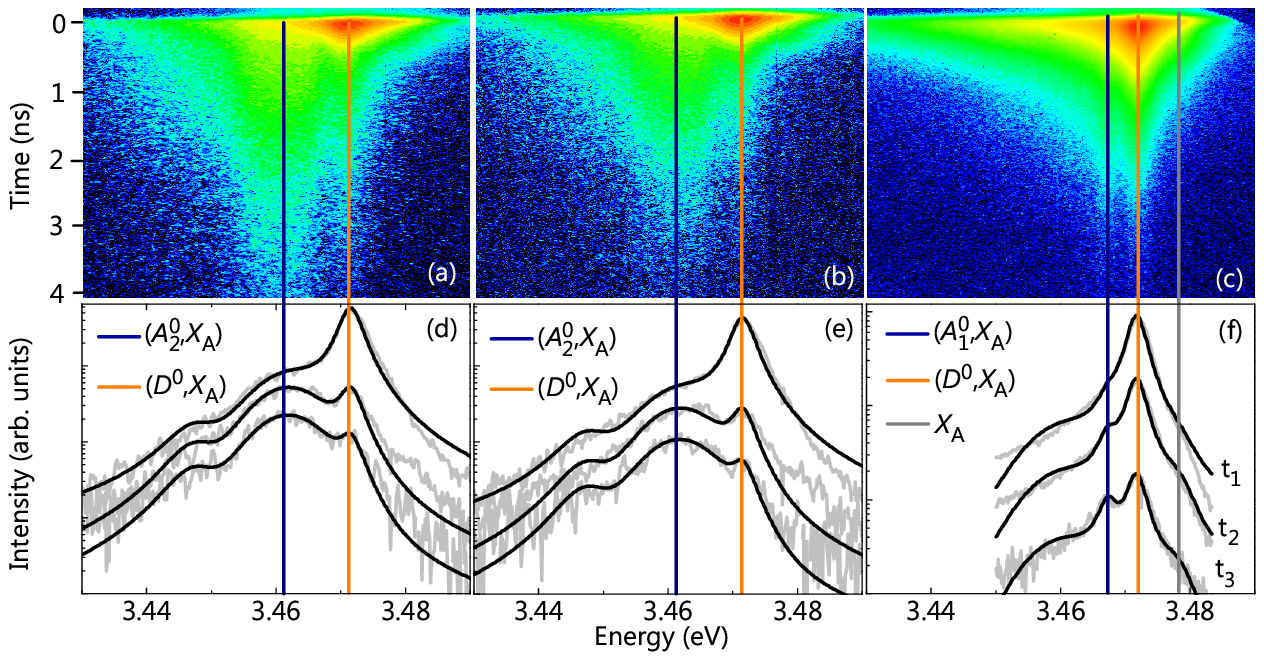} 
\caption{\label{fig:sc}(Color online) Streak camera image and transient PL spectra of sample A $\left[\text{(a) and (d)}\right]$, sample B $\left[\text{(b) and (e)}\right]$, and sample C $\left[\text{(c) and (f)}\right]$, respectively. The intensity in the streak camera images (a)--(c) is displayed on a logarithmic scale from blue (low intensity) to red (high intensity). The spectra [grey lines in (d)--(f)] are extracted from these images at times $t_1$ = 0.18, $t_2$ = 0.5, and $t_3$ = 1.35\,ns after excitation and are also displayed on a logarithmic intensity scale. Lineshape fits (black lines) to the experimental data allow us to perform a spectral deconvolution of the transitions (the $X_\text{A}$ transition can be reliably fit only for sample C, for which its intensity is comparatively high). The vertical lines represent the spectral positions of the individual transitions determined from the PL measurements presented in Fig.~\ref{fig:transients}.}
\end{figure*}
Figures~\ref{fig:transients}(a)--\ref{fig:transients}(c) show the PL spectra of the three samples on a logarithmic intensity scale. The dominant transitions in all spectra originate from the recombination of A excitons bound to neutral O and Si donors at $(3.4713 \pm 0.0001)$ [(O$^0,X_{\text{A}}$)] and $(3.4721 \pm 0.0001)$\,eV [(Si$^0,X_{\text{A}}$)], respectively. These values are essentially equal to those obtained in free-standing GaN layers within our experimental uncertainty.\cite{Freitas2002,Monemar2008} As expected for the comparatively large NW diameters, we do not observe a contribution from excitons bound to surface donors.\cite{Brandt_prb_2010} The observed linewidth of about 1\,meV for both transitions is thus determined by the residual microstrain within the GaN NWs.\cite{Kaganer2012} In addition to these dominant ($D^0,X_{\text{A}}$) transitions, all three samples exhibit a narrow line at 3.467\,eV stemming from the recombination of A excitons bound to neutral acceptors [($A^0_1,X_{\text{A}}$)].\cite{Monemar2008,Morkoc2008} Samples A and B exhibit an extra set of lines between 3.455 and 3.463\,eV [($A^0_2,X_{\text{A}}$)], which we attribute to the deeper acceptor states identified recently.\cite{Monemar2006, Monemar2008} Finally, a transition due to the recombination of B excitons bound to neutral donors [($D^0,X_\text{B}$)] at 3.475\,eV and from free A excitons ($X_\text{A}$) at 3.478\,eV is observed in all samples.

Figure~\ref{fig:transients}(d)--\ref{fig:transients}(f) displays the PL transients of the three samples integrated over a spectral window of 5\,meV width centered at the ($D^0,X_{\text{A}}$) transition energy. The decay is biexponential and remains virtually unchanged when varying the width of the spectral window between 2 and 20\,meV. The two components of the transients differ significantly in their decay time, particularly for samples A and B. The integrated intensity  is dominated by the short component, accounting for 85\%, 90\%, and 85\% for samples A, B, and C, respectively. The biexponential decay thus cannot be caused by the integration over the two transitions related to excitons bound to O and Si, since the intensity of these transitions is comparable [cf.\ Fig.~\ref{fig:transients}(a)--\ref{fig:transients}(c)]. Moreover, the lifetimes of excitons bound to O and Si were reported to be similar.\cite{Monemar2008,Monemar2010}

The biexponential decay can neither be attributed to nonradiative recombination of bound excitons in close proximity to the surface.\cite{Gorgis2012} Following Ref.~\onlinecite{Gorgis2012} and assuming surface recombination to be the dominant nonradiative decay channel for donor-bound excitons situated close to the surface, the short decay time of 90\,ps measured for samples A and B would correspond to an average NW diameter of 23\,nm, in blatant disagreement with the actual diameter distribution of the NW arrays under investigation [cf.\ insets of Fig.~\ref{fig:transients}(d)--\ref{fig:transients}(e)]. Moreover, to explain the amplitude of the short component would require 85\% to 90\% of \textit{all} donors to be in close proximity to the surface and even with the exact same distance. Besides the fact that this situation is of course entirely unlikely, it would manifest itself also in an energy shift of the transition,\cite{Brandt_prb_2010} which we do not observe in the PL spectra shown in Figs.~\ref{fig:transients}(a)--\ref{fig:transients}(c).

Having ruled out the two most obvious possibilities for a biexponential decay, we next examine the transient PL spectra of the samples. Figures~\ref{fig:sc}(a)--\ref{fig:sc}(c) show the raw streak camera images obtained after pulsed excitation. Immediately after excitation and up to a time of about 0.5\,ns, the ($D^0,X_{\text{A}}$) transition clearly dominates the spectra. For longer times, the ($A^0,X_{\text{A}}$) transition takes over as the dominant line in the spectra, i.\,e., its decay is significantly slower than that of the ($D^0,X_{\text{A}}$) transition. 

This result can be inspected more closely in Figs.~\ref{fig:sc}(d)--\ref{fig:sc}(f), which display transient spectra extracted from the streak camera images at three different times after excitation, namely, at  $t_1$ = 180,  $t_2$ = 500, and $t_3$ = 1350\,ps. Between $t_1$ and $t_2$, the ($D^0,X_{\text{A}}$) transition for samples A and B decreases in intensity by an order of magnitude with respect to the ($A^0_2,X_{\text{A}}$) transition. Between $t_2$ and $t_3$, however, the intensity ratio between these two transitions stays the same, i.~e., they decay with a common time constant for long times. For sample C [Fig.~\ref{fig:sc}(f)], we observe a qualitatively similar behavior, but the ($A^0_1,X_{\text{A}}$) transition becomes comparable in intensity with the ($D^0,X_{\text{A}}$) transition only at longer time ($>3$~ns).

The transient spectra shown in Figs.~\ref{fig:sc}(d)--\ref{fig:sc}(f) reveal a significant spectral overlap of the ($D^0,X_{\text{A}}$) and ($A^0,X_{\text{A}}$) lines. Even with the narrow spectral window used to obtain the transients shown in Figs.~\ref{fig:transients}(d)--\ref{fig:transients}(f), it is inevitable that we monitor a superposition of the corresponding transitions. Since the ($A^0,X_{\text{A}}$) transitions have a longer decay time than the ($D^0,X_{\text{A}}$) transition as seen in Fig.~\ref{fig:sc}, the biexponential decay may thus be interpreted as being simply due to the spectral overlap of these lines. The decay times of the two components of the transient would then correspond to the lifetime of the transition dominating the spectrum in a given time interval.   
  
\begin{figure*}[t]
\includegraphics*[width=15cm]{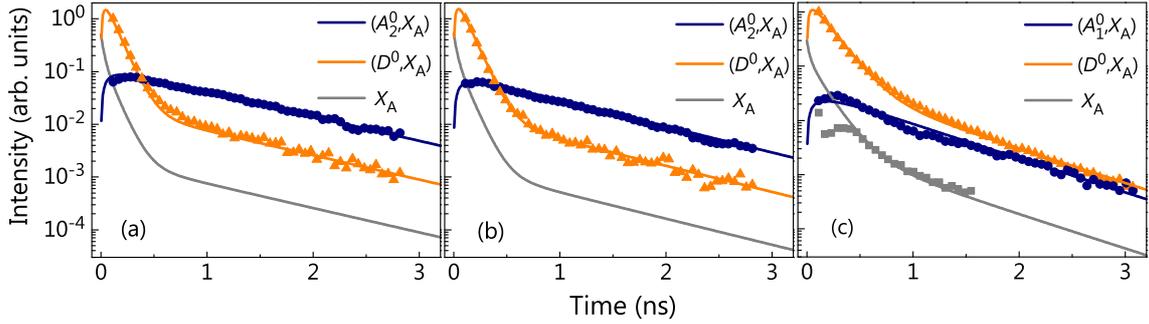} 
\caption{\label{fig:model}(Color online) PL transients for the ($D^0,X_{\text{A}}$) (triangles), ($A^0_1,X_{\text{A}}$) and ($A^0_2,X_{\text{A}}$) (circles), and $X_{\text{A}}$ [(squares, only in (c)] transitions obtained by the spectral deconvolution of the transient spectra [Fig.~\ref{fig:sc}(a)--\ref{fig:sc}(c)] for (a) sample A, (b) sample B, and (c) sample C. The solid lines represent the decay of these transitions as obtained by Eqs.~(\ref{model1})--(\ref{model3}). The fast initial decay ($50$\,ps) of the free exciton is caused by its capture by neutral donors and acceptors. Note the common decay time of all transitions at longer times which is a signature of their strong coupling.}
\end{figure*}

To examine this interpretation, we extract a series of transient spectra from the streak camera images and fit them by a sum of Voigt functions (three for samples A and B, four for sample C) as shown by the black lines in Figs.~\ref{fig:sc}(d)--\ref{fig:sc}(f). This spectral deconvolution allows us to explore the decay dynamics of each radiative recombination channel \textit{separately}. Figure \ref{fig:model} shows the time-dependent intensities of each transition as obtained by the deconvolution. While the ($A^0_1,X_{\text{A}}$) and ($A^0_2,X_{\text{A}}$) transients are monoexponential, the ($D^0,X_{\text{A}}$) transient is still clearly biexponential. This behavior is thus \emph{not} caused by the spectral overlap, and the above naive interpretation of the decay times of the two components of this transient is incorrect.

The key for the understanding of this result is the observation that the ($D^0,X_{\text{A}}$) and ($A^0,X_{\text{A}}$) transients are strictly parallel at long times. In addition, the $X_{\text{A}}$ and ($D^0,X_{\text{A}}$) transients for sample C are found to evolve in parallel, very similar to the results reported by Korona~\cite{Korona2002} for bulk GaN and Corfdir \textit{et al.}~\cite{Corfdir2009d} for GaN NWs. These transitions thus exhibit a common decay time, suggesting a strong coupling between all states participating in radiative recombination.\cite{Brandt1998b, Korona2002,Corfdir2009d} To facilitate a quantitative analysis of our data and to extract the actual lifetimes of these states, we model the time-dependent densities of the $X_{\text{A}}$ ($n_{\text{F}}$), ($D^0,X_{\text{A}}$) ($n_{\text{D}}$), and ($A^0,X_{\text{A}}$) ($n_{\text{A}}$) states by the following set of coupled rate-equations:
\begin{align}
\label{model1}
\frac{dn_{\text{F}}}{dt} & = -b_{\text{D}} n_{\text{F}} \left(N_{\text{D}}-n_{\text{D}}\right) -b_{\text{A}} n_{\text{F}} \left(N_{\text{A}}-n_{\text{A}}\right) \\ \nonumber &\quad +\hat{\gamma}_{\hspace{0.3mm}\text{D}}n_{\text{D}}+\hat{\gamma}_{\text{A}}n_{\text{A}} -\gamma_{\hspace{0.3mm}\text{F}}n_{\text{F}},\\
\label{model2}
\frac{dn_{\text{D}}}{dt} & = b_{\text{D}} n_{\text{F}} \left(N_{\text{D}}-n_{\text{D}}\right) - \hat{\gamma}_{\hspace{0.3mm}\text{D}}n_{\text{D}} -\gamma_{\hspace{0.3mm}\text{D}}n_{\text{D}}, \\
\label{model3}
\frac{dn_{\text{A}}}{dt} & = b_{\text{A}} n_{\text{F}} \left(N_{\text{A}}-n_{\text{A}}\right) - \hat{\gamma}_{\text{A}}n_{\text{A}} -\gamma_{\text{A}}n_{\text{A}},
\end{align}
with the initial densities $n_{\text{F}}(0)=n_{\text{F}}^0$, and  $n_{\text{D}}(0)= n_{\text{A}}(0)=0$.
\begin{figure}[b]
\includegraphics*[width=7cm]{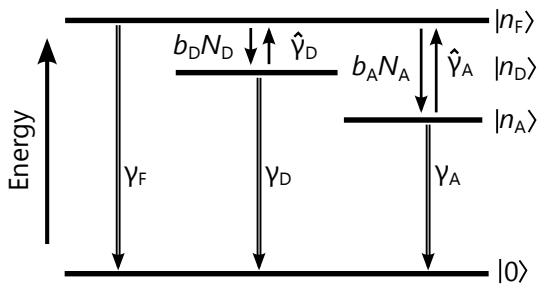} 
\caption{\label{fig:ediagram}Schematic energy diagram visualizing  Eqs.~(\ref{model1})--(\ref{model3}). The involved states are denoted by $| n_i \rangle$, and the crystal groundstate is represented by $| 0 \rangle$.}
\end{figure}
The first terms of Eqs.~(\ref{model1})--(\ref{model3}), which are illustrated in the scheme displayed in Fig.~\ref{fig:ediagram}, describe the capture of free excitons by neutral donors and acceptors with a total density $N_{\text{D}}$ and $N_{\text{A}}$ and the rate coefficients $b_{\text{D}}$ and $b_{\text{A}}$, respectively. The second terms account for the dissociation of the bound excitons with the rate constants $\hat{\gamma}_{\hspace{0.3mm}\text{D}}$ and $\hat{\gamma}_{\hspace{0.3mm}\text{A}}$ and the third ones for the recombination of free and bound excitons with the rate constants $\gamma_{\hspace{0.3mm}\text{F}}$,  $\gamma_{\hspace{0.3mm}\text{D}}$, and  $\gamma_{\hspace{0.3mm}\text{A}}$. These rate constants are the inverse of the effective decay times measured experimentally and implicitly contain radiative ($\gamma_{i,\text{r}}$) and nonradiative ($\gamma_{i,\text{nr}}$) contributions. The PL intensity of each transition is then given by $\gamma_{i,\text{r}}\,n_{i}$.\cite{*[{The radiative rate constant $\gamma_{i,\text{r}}$ directly determines the peak intensity of the transient [see }] [{]. To reproduce the experimentally observed peak intensities of the different transitions for each sample, we assume a radiative lifetime for the ($D^0,X_{\text{A}}$) of $\gamma^{-1}_{i,\text{r}}=1$\,ns (see Refs.~11 and 12) which in turn sets the radiative lifetimes for the $X_{\text{A}}$, ($A^0_1,X_{\text{A}}$), and ($A^0_2,X_{\text{A}}$) transitions  to 10, 7.7, and 5.5\,ns, respectively.}] Brandt2002} The free parameters of our model are the rate constants $\gamma_i$ for recombination, $\hat{\gamma}_i$ for the dissociation of bound excitons, and the rate constants for the capture of free excitons ($b_i N_i$).\footnote{Due to our low excitation density, we remain in the small-signal regime such that $N_i \gg n_i$. Thus, the product $b_i N_i$ approximates the capture dynamics of free excitons very well.} The solid lines in Fig.~\ref{fig:model} depict the simulated PL transients based on a numerical solution of Eqs.~(\ref{model1})--(\ref{model3}) using values for the free parameters as summarized in Tab.~\ref{tab:parameters}. The obtained capture rate constants are consistent with the experimentally observed rise times of the respective PL lines (not shown here).
\begin{table}[b]
\caption{Summary of the free parameters, all in units of ns$^{-1}$, of the rate-equation model [Eqs.~(\ref{model1})--(\ref{model3})] used for computing the PL transients shown in Fig.~\ref{fig:model}.}
\begin{ruledtabular}
\begin{tabular}{lccccccc}
Sample & $\gamma_{\hspace{0.3mm}\text{F}}$ & $\gamma_{\hspace{0.3mm}\text{D}}$ & $\gamma_{\hspace{0.3mm}\text{A}}$ & $\hat{\gamma}_{\hspace{0.3mm}\text{D}}$ & $\hat{\gamma}_{\text{A}}$ & $b_{\text{D}}N_{\text{D}}$ & $b_{\text{A}}N_{\text{A}}$ \\
\colrule 
A  & 8 & 11  & 0.5 & 10 & 0.65 & 20 & 2.8 \\
B  & 8 & 11  & 0.6 & 10 & 0.60 & 20 & 2.0 \\
C  & 3 & 7.5   & 0.4 & 10  & 1.3 & 26 & 2.8 \\
\end{tabular}
\end{ruledtabular}
\label{tab:parameters}
\end{table}

The excitonic states can be depopulated not only by recombination, but also by dissociation as depicted in Fig.~\ref{fig:ediagram}. The experimentally observed decay times are thus not necessarily equal to the actual lifetimes of these states. In this respect, our simulations provide a valuable guide for the interpretation of the experimentally observed transients. With the parameters listed in Tab.~\ref{tab:parameters}, the fast component of the biexponential decay of the ($D^0,X_{\text{A}}$) transition is essentially given by its effective lifetime $1/\gamma_{\hspace{0.3mm}\text{D}}$ and is thus governed by nonradiative recombination of the ($D^0,X_{\text{A}}$) complex. In contrast, the slow component is caused by a re-population of the ($D^0,X_{\text{A}}$) state due to its coupling with the deeper acceptor-bound excitons. In this particular case, its decay rate is approximately given by  $\gamma_{\text{A}}+\hat{\gamma}_{\text{A}}$ and thus results from the simultaneous dissociation and recombination of the ($A^0,X_{\text{A}}$) complex. 

At first glance, the strong coupling of the exciton states suggested by our results is surprising given the low measurement temperature of 10\,K. Corfdir \textit{et al.}~\cite{Corfdir2009d} attributed the parallel temporal evolution of the $X_{\text{A}}$ and ($D^0,X_{\text{A}}$) states at a lattice temperature of 8\,K to an enhanced thermal dissociation of bound excitons due to an electronic (carrier) temperature of 35\,K deduced from the high-energy tail of the transient spectra. Despite the low excitation density used in the present experiments, we obtain similar values from the exponential high-energy tail of the transient PL spectra immediately after excitation. However, for an electronic temperature of 35\,K and an exciton binding energy of 6--7\,meV, detailed balance arguments would predict a significantly smaller ratio of dissociation and capture rate constants than that obtained from the fits.\cite{Brandt1998b} 

We propose that the enhanced dissociation rate of bound excitons evident from our experiments is non-thermal in nature and related to the presence of electric fields within the GaN NWs.\cite{Calarco2011} The strength of these fields, which arise from the pinning of the Fermi level at the NW sidewall \emph{M}-plane surfaces,\cite{Calarco2005, VandeWalle2007} amounts to 10 to 17\,kV/cm for a moderate doping density of $2 \times 10^{16}$\,cm$^{-3}$ and the present range of NW diameters.\cite{Pfuller2010} Fields of this magnitude are theoretically expected to directly ionize the ($D^0,X_{\text{A}}$) complex\cite{Blossey1971,Yamabe1977a,Pedros2007} and have been experimentally found to quench the ($D^0,X_{\text{A}}$) line in GaN layers due to the dissociation of donor-bound excitons by impact ionization.\cite{Pedros2007} Note that the magnitude of these fields depends linearly on NW diameter and doping concentration for the characteristic dimensions of GaN NWs. For the same doping level, these fields are thus significantly weaker in thin GaN NWs such as investigated in Ref.~\onlinecite{Gorgis2012}. However, since they are an inherent property of GaN NWs of small to medium diameter, their effect on the exciton dynamics in these NWs must not be ignored.

Finally, our results imply that the lifetime of the ($D^0,X_{\text{A}}$) complex in thick GaN NWs is short and governed by a nonradiative decay channel not related to the NW surface. The actual origin of this decay channel is currently under investigation and will be the subject of a forthcoming publication. At present, we can firmly state that the nonradiative process is not intrinsic to GaN NWs in that it is neither related to the free surface nor to an excitonic Auger\cite{Kharchenko1990} process. In particular with regard to the latter, an increase of the fluence of the excitation by one order of magnitude results in a clear increase of the decay time, i.~e., the nonradiative process can be saturated. Schlager \emph{et al.}\cite{Schlager2011} even observed lifetimes up to 1\,ns (i.~e., close to the radiative one) by exciting very thick GaN NWs with a fluence two orders of magnitude larger than that used in the present work. For small-signal excitation as in the present work, however, the internal quantum efficiency of the GaN NWs under investigation is not larger than 20\% even at 10\,K. Whether higher values can be achieved in a different growth regime remains to be seen.

The authors would like to thank Vladimir Kaganer for providing a robust and fast batch fitting routine, the AMO GmbH for the preparation of the pre-patterned substrates and Manfred Ramsteiner for a critical reading of the manuscript. The work was partially funded by the German BMBF joint research project MONALISA (Contract No.~01BL0810).

\end{document}